\documentstyle[12pt]{article}
\begin{document}
\parskip=6pt
\baselineskip=22pt
{\raggedleft{\sf${ ASITP}$-94-25\\}}
\bigskip\bigskip\bigskip\bigskip\bigskip
\medskip
\centerline{\Large\bf A $SU(2)$ Generalized Gauge Field Model}
\vspace{2ex}
\centerline{\Large\bf With Higgs Mechanism
\footnote{\sf Work
supported in part by
The National Natural Science Foundation of China.}}
\vspace{20mm}
\centerline{\large \sf Han-Ying Guo ~and~ Jian-Ming Li  }
\vspace{3.5ex}
\centerline{\sf  CCAST (World Laboratory), P. O. Box 8730, Beijing 100080,
China;}
\vspace{1.5ex}
\centerline{\sf  Institute of Theoretical Physics, Academia Sinica,
P. O. Box 2735, Beijing 100080, China.\footnote{\sf Mailing address.}\\}
\vspace{8ex}

\centerline{\large \sf Abstract}

\bigskip
{\it By means of the non-commutative differential geometry, we construct
an $SU(2)$ generalized gauge field model. It is of
$SU(2) \times \pi_4(SU(2))$ gauge invariance. We show that
this model not only includes the Higgs field automatically on the equal footing
with ordinary Yang-Mills gauge potentials  but also is stable against
quantum correlation.}\\

\vspace{5mm}

\vfill
\newpage

Unlike Yang-Mills gauge
fields,  Higgs fields and  Yukawa
couplings  seem to be
artificial although they play a very
important role in modern QFT.
Eventually, the price paid for them is the beauty of the gauge
principle.

Recently, we have generalized  the ordinary
Yang-Mills gauge theory in order to take both Lie groups and discrete
groups as gauge groups [1,2] and completed an approach to this generalized
gauge theory coupled to the fermions in the spirit of non-commutative
geometry [3, 4]. We have shown that  Higgs fields are
such  gauge fields  with respect to discrete gauge symmetry over 4-dimensional
space-time $M^4$ and the Yukawa couplings between Higgs and
fermions may automatically be introduced via generalized covariant derivatives.
In this approach, Higgs appears as discrete group gauge fields on the same
footing with ordinary Yang-Mills fields over spacetime $M^4$.  In other wards,
the beauty of the gauge principle may be regained. Of course,
how to understand the physical meaning of the discrete group to be gauged is a
most crucial point in this approach. On the other hand, like other approaches
[3-10] based upon the non-commutative differential geometry do not survive
the standard quantum correlation [11], the approach in [1,2] may also be
unstable against the standard quantum correlation unless there is certain
special mechanism to guarantee its stability.

In this letter, we will present an $SU(2)$ generalized gauge field model with
the Higgs mechanism and show that it will be able to get rid of all those
problems. The key point is that we take into account the fourth homotopy group
of $SU(2)$ as a discrete gauge group on the footing with the Yang-Mills gauge
group $SU(2)$.
It is well known that the fourth homotopy group of $SU(2)$ is non-trivial,
$\pi_4(SU(2))=Z_2$, i.e. the gauge transformations of
$SU(2)$ may be divided into two different equivalence classes [12].
Once the Yang-Mills fields for the gauge group
$SU(2)$ is introduced, the role played by its fourth homotopy
group must be taken into account. In view of the generalized Yang-Mills gauge
theory [1] based upon the non-commutative differential geometry, we should
also introduce the generalized gauge field with respect to this internal
discrete group $\pi_4(SU(2))$ due to the fact that the gauge transformations
depend on its elements.

Let the elements of $\pi_4(SU(2))=Z_2=\{ e, r\}$ be $\{ U_e, U_r \}$
where $ U_e$ represents the equivalence
class of the  topologically trivial gauge transformations  and $U_r$
the  topologically non-trivial class modulo topologically trivial gauge
transformations.
The model under construction not only includes leptons $\psi(x,h),
h\in Z_2$, $SU(2)$ Yang-Mills gauge potentials $A_{\mu}(x,h)$
and Higgs $\Phi(x,h)$ with respect to this $Z_2$, but also combines both the
Yang-Mills gauge potential and the
Higgs on the equal footing as a generalized Yang-Mills gauge potential.

Let us regard those fields as elements of function space on $M^4$ as well as
on $SU(2) \times  \pi_4(SU(2))$ and assign them into two sectors
according to two elements of $ \pi_4(SU(2))=Z_2=\{ e, r \}$ as follows:
\begin{equation}\begin{array}{cl}
\psi(x,e)=\psi(x)=\left(
\begin{array}{cl}
L\\
R\end{array}\right); ~~\psi(x,r)=\psi^r(x)=\left(
\begin{array}{cl}
L^r\\
R\end{array}\right)\\[6mm]
A_{\mu}(x,e)=A_{\mu}(x)=\left(
\begin{array}{ccc}
L_{\mu}&0\\[1mm]
0&0\end{array}\right); ~~A_{\mu}(x,r)=A^r_{\mu}(x)=\left(
\begin{array}{ccc}
L^r_{\mu}&0\\[1mm]
0&0\end{array}\right)\\[6mm]
\Phi(x,e)=\Phi(x)=\left(
\begin{array}{ccc}
{\frac{\mu}{\lambda}}&{-\phi}\\
{-\phi^r}^{\dag}&{\frac{\mu}{\lambda}}\end{array}\right);~~\Phi(x,r)=\Phi^r(x);
\end{array}\end{equation}
with  $L^r=UL$, $\phi^{r}=U\phi$, $UU\dag=1$, $U$ is
a topologically non-trivial $SU(2)$ gauge transformation.
In (1), $L ~(R)$ is the left (right) handed fermion which is an $SU(2)$ doublet
(singlet), $L_{\mu}=-ig\frac{\tau_i}{2}W^{i}_{\mu}$ the $SU(2)$-gauge
potential, $\phi$ also an $SU(2)$ doublet,
$\mu$ and $\lambda$ two constants.

It should be mentioned, however, that the assignments (1)
not only assign the fields to the elements of $Z_2$ but also
imply that all fields are  arranged
into certain matrices. In fact, this aspect of the arrangements is nothing to
do with discrete gauge symmetry but  for convenience in the forthcoming
calculation. Of course, it must be kept in
mind that this is a working hypothesis and sometimes one should
 avoid certain extra
constraints coming from this working hypothesis.

{}From the general framework in [1], it follows the generalized connection
one-form
\begin{equation}
A(x,h)=A_{\mu}(x,h)dx^{\mu}+\frac{\lambda}{\mu}\Phi(x,h)
\chi,~ ~~h\in Z_2,
\end{equation}
where $\chi$ denotes ${\chi}^{r}$, a one form on the function space on
$\pi_4(SU(2))$,
and the generalized curvature two-form
\begin{equation}
\begin{array}{cl}F(h)
&=dA(h)+A(h)\otimes A(h) \\[4mm]
&=\frac{1}{2}F_{\mu\nu}(h)
{dx}^{\mu}\wedge{dx}^{\nu}+
\frac{\lambda}{\mu} F_{\mu r}(h){dx}^{\mu}\otimes{\chi}
+\frac{{\lambda}^{2}}{{\mu}^{2}}
 F_{rr}(h){\chi}\otimes{\chi}.\end{array}
\end{equation}
Using the above assignments, we get
\begin{equation}\begin{array}{cl}
F(x,e)
&=F^r(x,r)\\[4mm]
&=\frac{1}{2}\left(
\begin{array}{cc}
L_{\mu\nu}&0\\[1mm]
0&0\end{array}\right){dx}^{\mu}\wedge{dx}^{\nu}
+\frac{\lambda}{\mu}
\left(
\begin{array}{cc}
0&{-D_{\mu}\phi}\\
-(D_{\mu}\phi^{\dag})^r&0\end{array}\right)
{dx}^{\mu}\otimes{\chi}\\
&+\frac{{\lambda}^{2}}{{\mu}^{2}}
\left(
\begin{array}{cc}
{\phi{\phi}^{\dag}-\frac{{\mu}^2}{\lambda^2}}&0\\
0&{\phi^r}^{\dag}{\phi}^r-\frac{\mu^2}{\lambda^2}\end{array}\right)
{\chi}\otimes{\chi};\\[6mm]
\end{array}\end{equation}
where $L_{\mu\nu}=-ig\frac{\tau_i}{2}W^i_{\mu\nu}, ~~
D_{\mu}{\phi}={\partial}_{\mu}{\phi}+L_{\mu}{\phi}
=({\partial}_{\mu}-ig\frac{{\tau}_{i}}{2}W^{i}_{\mu})\phi.$

Having these building blocks,
we may  introduce the generalized gauge invariant
Lagrangian with respect to each element of $Z_2$, then take the Haar
integral of them over $Z_2$ to get the entire Lagrangian of the model.
Under certain consideration on the normalization in the Lagrangian, we may get
a Lagrangian without any extra constraints among the coupling constants and the
mass parameters at the tree level.

For the Lagrangian of the bosonic sector, we have
\begin{equation}\begin{array}{cl}
{\cal {L}}_{YM-H}(x,e)
&={\cal {L}}^r_{YM-H}(x,r)\\[4mm]
&=-\frac{1}{4N_L}Tr(L_{\mu\nu}L^{\mu\nu})\\[4mm]
&~+\frac{2}{N}\eta\frac{{\lambda}^{2}}{{\mu}^{2}}
Tr(D_{\mu}\phi(x))(D^{\mu}\phi(x))^{\dag}\\[4mm]
&~-\frac{2}{N}\eta^2\frac{{\lambda}^{4}}{{\mu}^{4}}
Tr(\phi(x){\phi(x)}^{\dag}-\frac{{\mu}^{2}}
{{\lambda}^{2}})^{2}+const;
\end{array}\end{equation}
where  $N_L$ and $N$ are
normalization constants introduced here to avoid
some extra constraints from the matrix arrangement in (1), $\eta$
a metric parameter defined by
$ \eta=< \chi, \chi >, ~Dim(\eta)={\mu}^2$.
The normalization of the coefficients of each term  results
\begin{equation}
N_L=\frac{g^{2}}{2},  ~N=2\frac{{\lambda}^2}{{\mu}^2}\eta.
\end{equation}

For the fermionic sector, the Lagrangian  may also be given as follows:
\begin{equation}\begin{array}{cl}
{\cal{L}}_{F}(x,e)
&={\cal{L}}^r_{F}(x,r)\\[4mm]
&=i\overline{L}\gamma^{\mu}({\partial}_\mu+L_\mu)L
+i\overline{R}\gamma^{\mu}{\partial}_{\mu}R
+\lambda (\overline{L}\phi{R}+\overline{R}{\phi}^{\dag}L).
\end{array}\end{equation}

Then the entire Lagrangian for the model reads:
\begin{equation}
{\cal{L}}(x)
=\frac{1}{2}\sum_{h=e, r}\{{\cal{L}}_{F}(x,h)+{\cal {L}}_{YM-H}(x,h)\}.
\end{equation}

It is remarkable that  this is a Lagrangian with the Higgs mechanism of
spontaneously symmetry breaking type and the Yukawa couplings included
automatically. In fact, the Higgs potential takes its minimum value at
 $Tr(\phi{\phi}^{\dag})=(\frac{\mu}{\lambda})^{2}$ and
the continuous gauge symmetry $SU(2)$ will spontaneously be broken down when
 the vacuum expectation value is taken as, say,
\begin{equation}
<{\phi}>=\left(
\begin{array}{cl}
0\\[1mm]
\frac{\rho_0}{\sqrt{2}}
\end{array}\right),
\end{equation}
where $\rho_0=\sqrt{2}\frac{\mu}{\lambda}$.
Now we take the vacuum expectation value of $\phi$ and introduce a new field
$\rho(x)$
\begin{equation}
{\phi}=\left(
\begin{array}{cl}
0\\[1mm]
\frac{\rho_0+\rho(x)}{\sqrt{2}}\end{array}\right).
\end{equation}
Then we have in the Lagrangian of the bosonic part
\begin{equation}\begin{array}{cl}
&Tr\{D_{\mu}\phi(D_{\mu}\phi)^{\dag}-
\eta\frac{\lambda^2}{\mu^2}(\phi{\phi}^{\dag}
-\frac{{\mu}^{2}}{{\lambda}^{2}})^{2}\}\\[4mm]
&=\frac{1}{2}{\partial}_{\mu}\rho{\partial}^{\mu}\rho+\frac{g^{2}}{4}
(\rho_0+\rho)^{2}W_{\mu}^{-}W_{\mu}^{+}+\frac{1}{8}g^{2}
(\rho_0+\rho)^{2}W^3_{\mu}W^3_{\mu}\\[4mm]
&~~~-\eta\frac{\lambda^2}{\mu^2}{\rho}^{2}(\rho_0^{2}+\rho_0\rho
+\frac{{\rho}^{2}}{4})+const.
\end{array}\end{equation}
It is easy to see that only all gauge bosons
$W^{\pm}$ and $W^3$  and Higgs $\rho$ become massive:
\begin{equation}
M_{W}=\frac{1}{2}g\rho_0,~~~
M_{Higgs}=2\sqrt\eta.
\end{equation}
While for the fermions, one of the components of $L$, the down fermion in the
$SU(2)$-doublet, becomes massive with mass
$\mu$ and others
remain massless. If  the metric parameter $\eta$ is free of choice, there do
not exist any constraints among the coupling constants.
This is different from other approaches [5-10].

Let us now  summarize what we have done.
Based on a first principle, the generalized gauge principle,
we have constructed an $SU(2)$ generalized gauge field model with
$\pi_4(SU(2))=Z_2$ taken as discrete gauge symmetry.
The Higgs mechanism is automatically included in this generalized gauge
theory model.

It is worthy to point out that there are several advantages in this approach.
First, this $\pi_4(SU(2))$ is a most natural and meaningful internal
symmetry to be gauged in the model. What we have
done here is just to combine the ordinary Yang-Mills gauge theory with the
non-commutative differential calculus in the function
space on this discrete group to formulate a generalized gauge theory with Higgs
and spontaneously symmetry breaking. In other wards, the Higgs mechanism should
be introduced automatically on the equal footing with ordinary Yang-Mills gauge
fields, if the role played by the fourth homotopy group of the gauge
groups would be taken into account together with the gauge groups themselves
at very beginning.

It is even more remarkable that the
approach presented here is stable against quantum correlation.
One of the reasons is that there are no extra
constraints among the parameters at
the tree level so that we do not need to pay attention to them in the course of
quantization.
 Another reason may be more essential. Namely,
since the Higgs potential is automatically introduced,
 the $SU(2)$ gauge symmetry should be
spontaneously broken down at the tree level  in this model. Therefore, this
$\pi_4(SU(2))=Z_2$ symmetry is also broken down
as long as the VEV for the Higgs is taken. Consequently, what we got is
the same version as an
ordinary $SU(2)$ Yang-Mills model with Higgs mechanism and of course we do not
need to concern about this $\pi_4(SU(2))=Z_2$-gauge symmetry when we consider
the quantum correlation in
the model. Needless to say, this very important point is completely different
from other approaches to the Higgs by means of the
non-commutative differential geometry. In fact, Connes like approaches [5-10]
do not survive the standard quantum correlation [11].

It is clear that the model presented here is not phenomenologically realistic
but it  can be generalized to other gauge theory models, such as
the Weinberg-Salam model and the standard model for the
electroweak-strong interaction. In those cases, this approach may also shed the
light on the Higgs pattern. Since $\pi_4(SU(N))=0, ~N\not=2$, Higgs mechanism
of
this type should not appear in the gauge field sectors of $SU(N), ~N\not=2$.
On the other hand, the model  may also be
generalized to
the case of $SU(2)_L\times SU(2)_R$ gauge invariance with
$ \pi_4(SU(2)_L \times SU(2)_R)=Z_2 \times Z'_2$ and it may be applied to
the left-right symmetric model. Furthermore, since the fourth homotopy
group of $SU(5)/ ( SU(3) \times SU(2) \times U(1))$ is also non-trivial, it may
play certain role in the SU(5)-GUT together with
$\pi_4( SU(3) \times SU(2) \times U(1))=Z_2$. And all models of
this kind may have the same advantages as the one presented in this letter.
Especially, all of them should also be stable against quantum correlation.

\bigskip
\bigskip

\newpage

\bigskip

\begin{enumerate}

\item H.G. Ding, H.Y. Guo, J.M. Li and K. Wu,
Comm. Theor. Phys. {\bf 21} (1994) 85-94;

Higgs as gauge  fields on discrete
groups and standard models for electroweak and electroweak-strong interactions;
To appear in Z.Phsik. {\bf C}.

\item
H.G. Ding, H.Y. Guo, J.M. Li and K. Wu,
 J. Phys.  {\bf A 27} (1994) L75-L79;  ibid. L231-L236.

\item A. Connes, \-{\it Non-Commutative \- Geometry}~~
\-English \- translation  of \- Geometrie \- Non-Commutative, IHES \- Paris,
 Interedition.

 \item
 A. Sitarz,  Non-commutative Geometry and  Gauge Theo\-ry on Dis\-crete
Groups, preprint {\bf TPJU}-7/1992.

\item A. Connes, in: The Interface of  Mathematics and Particle Physics,\\
eds. D. Quillen, \- G. Segal and S. Tsou \- (Oxford U. P, Oxford 1990).

\item A. Connes and J. Lott, \- Nucl. Phys. (Proc. Suppl.) {\bf B18}, 44
(1990).

\item A. Connes and Lott, \- Proceedings of 1991 \-Cargese Summer \-
Conference.

\item D. Kastler, Marseille, CPT preprint {\bf CPT-91}/P.2610, {\bf
CPT-91}/P.2611.

\item A. H. Chamseddine, \- G Felder and J. Fr\"ohlich, \- Phys. Lett. \-
 {\bf 296B} (1993) 109, \- Zurich preprint  \- {\bf ZU-TH}-30/92 and
Zurich \- preprint \- {\bf ETH-TH}/92/44.
 A. H. Chamseddine \- and J. Fr\"ohlich, \- $SO(10)$ \- Unification in \-
 Noncommutative
\- Geometry, {\bf ZU-TH}-10/1993.

\item R. Coquereaux, G. Esposito-Far\'ese and G Vaillant, Nucl Phys
{\bf B353}  689 (1991).

\item E. \'Alvarez, J.M. Gracia-Bondia and C.P. Martin,
Phys. Lett. {\bf B306}, 53 (1993).

\item S.T. Hu, Homotopy Theory, Academic Press, New York, 1959.

\end{enumerate}

\end{document}